# Mechanical design and analysis for a low beta squeezed half-wave resonator*

HE Shou-Bo[1;2;1)] HE Yuan[1] ZHANG Sheng-Hu[1] YUE Wei-Ming[1] ZHANG Cong[1] WANG Zhi-Jun[1]
WANG Ruo-Xu[1] XU Meng-Xin[1,2] HUANG Shi-Chun[1,2] HUANG Yu-Lu[1,2] JIANG Tian-Cai[1,2]
WANG Feng-Feng[1] ZHANG Sheng-Xue[1] ZHAO Hong-Wei[1]

1 Institute of Modern Physics, Chinese Academy of Sciences, Lanzhou 730000, China
2 University of Chinese Academy of Sciences, Beijing 100049, China

**Abstract** A superconducting half-wave resonator (HWR) of frequency=162.5 MHz and β=0.09 has been developed at Institute of Modern Physics. Mechanical stability of the low beta HWR cavity is a big challenge in cavity design and optimization. The mechanical deformations of a radio frequency superconducting cavity could be a source of instability, both in continues wave(CW) operation or in pulsed mode. Generally, the lower beta cavities have stronger Lorentz force detuning than that of the higher beta cavities. In this paper, a basic design consideration in the stiffening structure for the detuning effect caused by helium pressure and Lorentz force has been presented. The mechanical modal analysis has been investigated with finite element method(FEM). Based on these considerations, a new stiffening structure has been promoted for the HWR cavity. The computation results concerning the frequency shift show that the low beta HWR cavity with new stiffening structure has low frequency sensitivity coefficient, Lorentz force detuning coefficient $K_L$ and stable mechanical property.

**PACS** 29 20.Ej

## 1 Introduction

Superconducting (SC) coaxial half-wave resonator (HWR) was first proposed and fabricated by Argonne National Laboratory over twenty years ago [1]. The HWR cavities show their optimum application range for frequencies of ~150–350 MHz and for beam velocities of ~0.06<β<0.4, where they can prevail in some aspects over other resonator geometries. Recently, various superconducting HWR structures, generally including squeezed type [2], cylinder type [3], taper type [4], are widely developed in several proposed high-intensity light ion linac projects. The CADS injector-II program in superconducting HWR development at Institute of Modern Physics(IMP) includes a 162.5MHz squeezed type cavity with β= 0.09 [5]. In an initial phase, one squeezed type HWR cavity and two solenoids will be supplied, and will need to be tested together as they would operate in a machine-type cryomodule. Figure 1 presents the HWR cavity together with its helium tank, the mechanical tuner and the main coupler, in the configuration that will be tested in the test cryomodule at IMP.

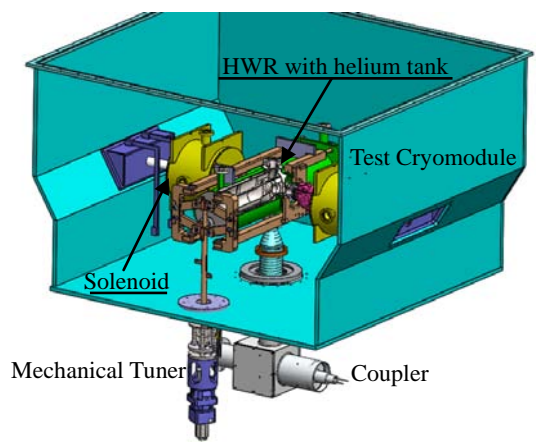

Figure 1. (colour online) Elevation view of β=0.09 squeezed type HWR cavity in test cryomodule.

*supported by Strategic Priority Research Program of CAS (XDA0302)and National Natural Science Foundation of China (91026004)
1)heshoubo@gmail.com

The SC HWR cavity was constructed from pure niobium with a specified pre-processed wall thickness of 2.8 mm. A 3 mm thick helium shell with attached 0.2 mm thick bellows, which exists between the helium tank and NiTi flange, allows axial movement to tune the HWR cavity.

Since SC the HWR cavity is highly sensitive to mechanical deformations due to its narrow bandwidth, evaluating the frequency shift and improving the cavity stability must be involved during the cavity prototype design. Usually, to reduce influences of mechanical deformations, such as helium pressure, Lorentz force detuning(LFD) and modal vibrations, it is possible to increase the wall thickness to enhance the cavity rigidity. However, this solution increases the cost and what is more important, it also reduces the efficiency of cooling by liquid helium. In order to improve the pressure sensitivity coefficient $df/dp$ and Lorentz force detuning coefficient $K_L$, the typical approach is to install a optimized stiffening structure on the cavity. In this paper, a basic consideration to optimize the stiffening structure is presented in the second section. Based on the pressure sensitivity coefficient, the stiffening structure design of the low beta squeeze type HWR cavity is introduced in the section III. General comparison between the stiffening design and previous design, including Lorentz force detuning effect and mechanical resonance, are discussed based on the calculating results with the finite element method software ANSYS multiphysics [6]. Conclusions and suggestions for future research are given in the last section.

## 2 Basic considerations

In this SC HWR cavity, the RF power produces radiation pressures on the inside cavity wall. Additionally, the liquid helium in the vessel also loads pressures on the outside cavity wall. The pressures deform the cavity wall, then causes the cavity resonant frequency shift. RF system needs to supply a surplus RF power to compensate the cavity frequency shift for keeping the cavity voltage constant $V_{acc}$ in a superconducting cavity with geometric shunt impedance and external quality factor, which is given by equation 1 (especially for over-coupled), as a conventional method.

$$P = \frac{V_{acc}^2}{4\frac{R}{Q}Q_{ext}}\left[\left(1+\frac{R}{Q}Q_{ext}\frac{I_b}{V_{acc}}\cos\phi_b\right)^2 + \left(2Q_{ext}\frac{\Delta f}{f_0}+\frac{R}{Q}Q_{ext}\frac{I_b}{V_{acc}}\sin\phi_b\right)^2\right] \quad (1)$$

Here $I_b$ is the average beam current, $\phi_b$ implies the accelerating phase, $f_0$ means the cavity frequency, and $\Delta f$ represents the cavity frequency shift. It is worth noting that higher levels of detuning require more RF power for SC HWR cavity, which will greatly add much to the cost of cavity operation.

The more effective method, which compensates the cavity frequency shift, is to minimize the frequency detuning and to employ a cavity tuner. The function of the stiffening structure is to control the frequency detuning and protect cavity. A good stiffening structure should: (i) have low helium pressure sensitivity; (ii) have effectively control the Lorentz force detuning; (iii) have low peak stress in the cavity; (iv) have low tuning sensitivity; (v) have safe mechanical resonance mode. The stiffening structure design for squeezed type HWR cavity follows these rules. The mechanical properties of niobium used during analysis in the following are listed in Table1.

Table 1. The mechanical properties of niobium during analysis.

| | |
|---|---|
| Poisson's ratio | 0.38 |
| Young's modulus/GPa | 105 |
| Yield strength[1)]/MPa | 50 |
| Yield strength[2)]/MPa | 140 |
| Tensile strength /MPa | 400 |

1) room temperature for 300K;   2) low temperature for 4.2K.

## 3 Cavity stiffening design

The low $\beta$ squeezed type HWR is characterized by a double-wall coaxial structure terminated by end cone, with squeezed middle section of outer conductor and integrated helium vessel, which makes itself very compact. From above, the stability of HWR cavity structure against any external distortions is the primer design goal. However, fluctuations in the helium bath pressure are most likely the main source of frequency detuning. Based on this consideration, a detailed mechanical design and analysis for the IMP squeezed type HWR cavity stiffening, which means minimizing the pressure sensitivity $df/dp$ as low as possible by optimization, are discussed in the following section.

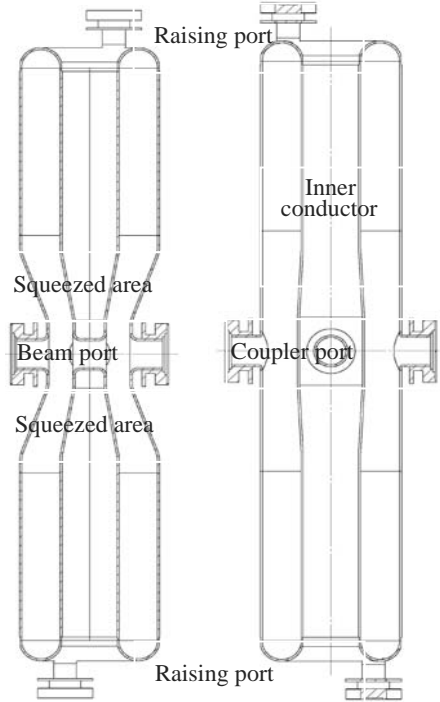

Figure 2. Section view of IMP squeezed type HWR cavity without stiffening structure.

The HWR cavity has a characteristic shape of squeeze in the middle of outer conductor as shown in Figure 2. The squeezed part of the cavity is a transition from a round shape to racetrack shape. According to the pressure vessel code and tuning study for this cavity [7], optimization should be given more consideration about structure stiffening in this region. From the Slater's Theorem[8], we can see that deformation in the electric field region will give an opposite contribution to the frequency shift, compared to the shift from deformation in the magnetic field region. Removing a small volume in the high magnetic field region will decrease the inductance, increasing the resonant frequency. Removing a small volume in the high electric field region will increase the capacitance, reducing the cavity frequency. The general basics of the cavity structural design are to avoid using the plane surfaces in the squeezed area as illustrated in Figure 2.

During the mechanical design, more than ten types of stiffening ribs have been analyzed to minimize the pressure sensitivity. Here simulation results of five major stiffening structures are introduced in detailed. For maximum accuracy, the complete cavity assembly was employed during the simulations. The tetrahedral elements with midside nodes are applied in the mesh of finite element solution. The same mesh is used for the entire simulation, switching the element types between structural and high frequency for different analyses, and updating the mesh after determining the deformation due to the pressure load rather than remeshing, which greatly improves analysis accuracy. Figure 3 presents the deformation results of HWR cavity with one atmosphere pressure load after adding stiffening structure.

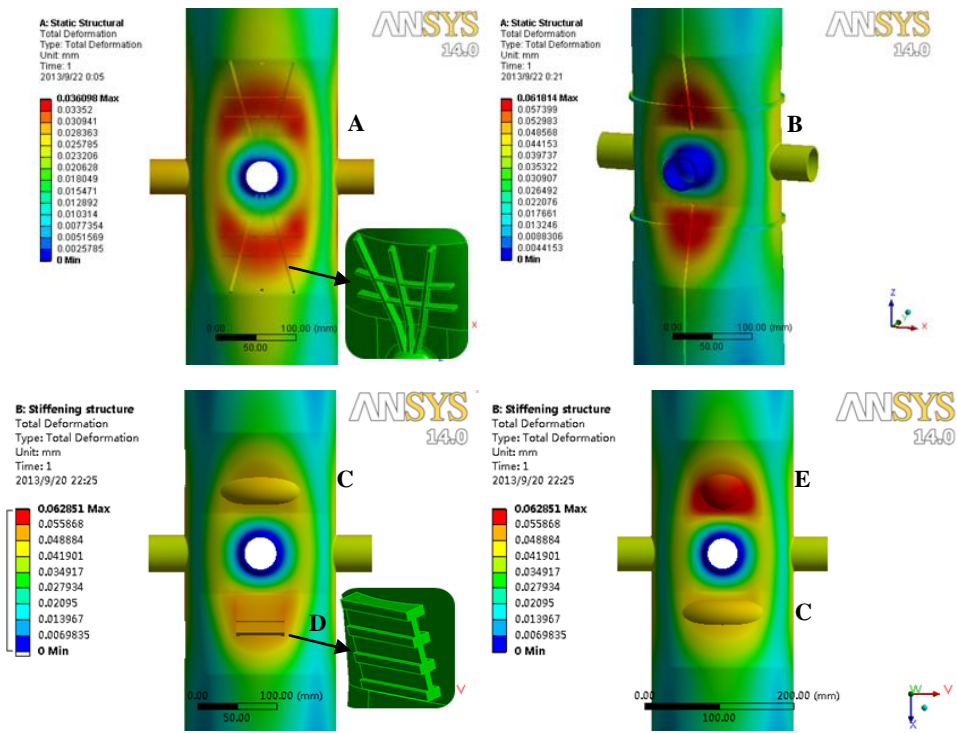

Figure 3. (colour online) The deformation results of HWR cavity with five major stiffening structures.

From Figure 3, the most sensitive region is located at the squeezed part. So during optimization, several stiffening structure, mainly focused on the squeezed area, have been simulated. Different stiffening ribs lead to diverse deformation in the outer conductor. Generally speaking, the deformation is proportional to the cavity frequency shift. Compared with the naked HWR cavity, the mechanical results of HWR with five stiffening structures are shown in the Table 2.

Table 2. The mechanical results of HWR with un-stiffening and stiffening under one atmosphere pressure load.

| Type No. | Max Disp./ $\mu m$ | Peak Stress/MPa | $\Delta f$ /KHz | $df/dp$ /(Hz/mbar) | Ratio[*] % |
| --- | --- | --- | --- | --- | --- |
| No stiffening | 93.6 | 39.9 | 17.9 | -17.7 | --- |
| Stiffening A | 36.1 | 25.9 | 12.1 | -11.9 | 32.8 |
| Stiffening B | 61.8 | 35.6 | 14.7 | -14.5 | 18.1 |
| Stiffening C | 51.3 | 38.2 | 13.6 | -13.4 | 24.3 |
| Stiffening D | 53.6 | 43.2 | 13.9 | -13.7 | 22.6 |
| Stiffening E | 62.9 | 44.7 | 15.8 | -15.6 | 11.9 |

*The ratio is a special value. It is defined by: change of pressure sensitivity for HWR cavity with each of stiffening structure divided by the one of naked HWR cavity without any stiffening structure, which can be presented by the equation: $\dfrac{\left| df/dp_{stiffening} - df/dp_{naked} \right|}{df/dp_{naked}}$.

As indicated in Table 2, simulations predicted the pressure sensitivity $df/dp$ to be 11.9 Hz/mbar for the stiffening A, which is significantly lower than the other stiffening structure. Simultaneously the peak stress on the cavity with stiffening ribs decreased greatly, which could prevent the HWR cavity from plastic deformation both at the room temperature and low temperature. Although the pressure sensitivity can be declined continuously by adding more complicated stiffening structures, a scheme of stiffening A in Figure 3 is adopted by IMP finally. The manufacture technology is taken into account, which is the main reason for selecting this scheme. Secondly, considering the cavity bandwidth of 235Hz, the pressure sensitivity of 11.9 Hz/mbar can meet the requirement of frequency tuning system. Recently the 162.5MHz squeezed type HWR cavity with the designed stiffening structure is welded and improved at the welding laboratory of Harbin Institute of Technology(HIT).

## 4 Mechanical resonance

Due to the high $Q_L$ of the HWR cavity, it is very sensitive to detuning. Mechanical resonances of cavity in a cryomodule can be excited by external vibrations, helium bath pressure fluctuations and etc., which would cause huge increases in the power required to keep the cavity at the design gradient. Stiffening ribs strongly affect the frequency of the mechanical resonances of the cavity. The three-dimensional finite element models in ANSYS, described in the previous section, were used to determine the mechanical resonant frequencies. Some illustrations of the first six modes in the HWR cavity are shown in Figure 4.

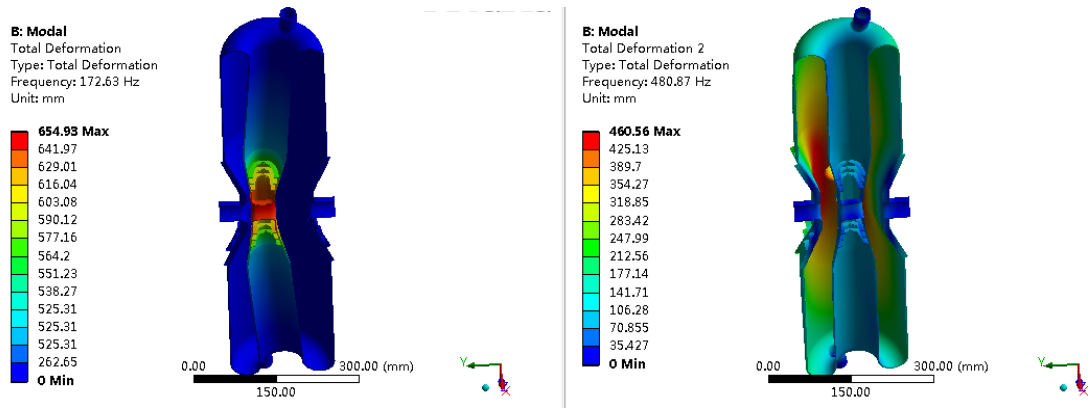

Figure 4. (colour online) Two vibration modes of HWR cavity with stiffening A type structure.

In modal analysis, the reported displacements do not reflect a true estimated value; rather they can indicate relative magnitude of a structure's response at a given frequency. Therefore, the modal results focus on the fundamental frequencies and associated mode shapes [9]. In Figure 4, the mode of vibration in left arises in the inner-conductor and the frequency of the mode is 172.6 Hz. For the outer-conductor vibration in right picture, the mode frequency is about 480.9 Hz. The frequencies of the first six modes in the IMP HWR cavity with stiffening structure and without stiffening structure are shown in Table 3.

Table 3. The first six modes of HWR cavity for two structures.

| Mode No. | Stiffening A/Hz | No stiffening/Hz |
| --- | --- | --- |
| Mode 1 | 172.63 | 168.67 |
| Mode 2 | 258.53 | 242.20 |
| Mode 3 | 480.87 | 410.44 |
| Mode 4 | 489.28 | 480.13 |
| Mode 5 | 559.56 | 505.42 |
| Mode 6 | 593.26 | 550.90 |

The simulation results of HWR cavity with stiffening A structure show that the mechanical mode with the lowest frequency is a longitudinal mode of 172.6 Hz, and the lowest frequency of the transverse mode is about 258.5 Hz. Normally, the lower the frequency of mechanical mode, the easier an environmental source that excites it can be found. Besides, noise intensity tends to increase as $1/f$ [10]. From above, it can be concluded that better stiffening ribs have mechanical modes at significantly higher frequencies, which is beneficial for two reasons. First, it makes active compensation of microphonics via the piezo tuner more effective. Secondly, the amplitudes of vibration sources tend to be lower at higher frequencies.

## 5   Lorentz force detuning

LFD was simulated in ANSYS, again using a procedure similar to the $df/dp$ simulations, but with a pressure load calculated from the Lorentz forces from fields in the cavity instead of a constant pressure. A special boundary condition is applied between the two beam ports for two cases, with stiffening and without stiffening structure. The simulation results of LFD coefficient are shown in Figure 5 for the two cases.

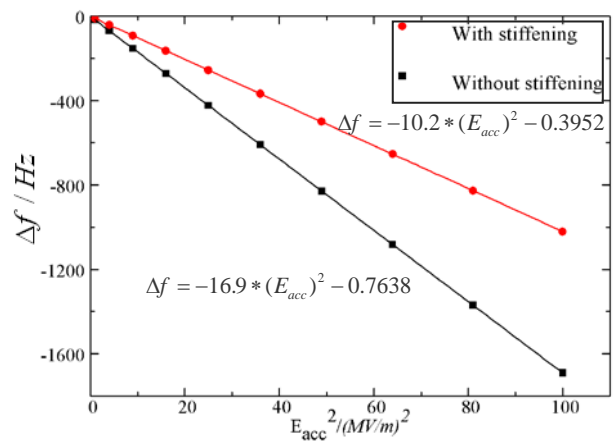

Figure 5. The fitting curve between frequency shift and accelerating gradient for two cases.

The boundary conditions strongly influence in the value for the Lorentz force coefficient. During the simulation, both the FEM models have been loaded the same boundary. As expected in Figure 6, the Lorentz force detuning coefficient for the HWR cavity with stiffening A structure has decreased to -10.2 $(MV/m)^2$ compared with coefficient for the naked cavity, -16.9 $(MV/m)^2$. Figure 6 illustrates the relationship between LFD and energy content for both stiffening structures. The stiffening A structure gives a low LFD coefficient after the mechanical design. From Figure 6, the frequency shift has been reduced to a greater extent by adding the new stiffening structure.

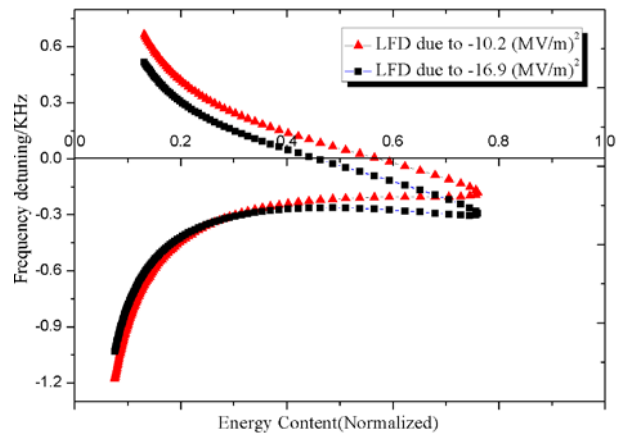

Figure 6. The relationship between LFD ( $E_{acc}$ =4.22MV/m) and energy content with two cases.

# 6  Conclusion

The superconducting RF linac of CADS injector-II requires operating the low beta HWR cavity at high loaded quality factor to reduce costs and maximize operational efficiency. Therefore the HWR cavity needs to have good mechanical stability and small sensitivity coefficient $df/dp$. The unstiffened cavity will be easier to manufacture than the stiffened cavity, but it will be more fragile during handling and its low frequency mechanical resonances will limit the bandwidth of the piezotuner. In this paper, several methods were presented to significantly reduce pressure sensitivity through optimization of the squeezed part of HWR cavity. For CW superconducting linac in IMP, a squeezed type HWR cavity is being fabricated to investigate the optimized stiffening structure at HIT. Considering the stiffening structure, the stable mechanical property, such as LFD and mechanical resonance, has been improved greatly. A novel taper type structure half-wave resonator should be promoted in the future for further increasing both mechanical stability and rf performance.